\documentclass[12pt]{article}
\usepackage{amssymb}
\usepackage{color,graphicx}
\usepackage{amsmath}

\newcommand{\be}{\begin{equation}}
\newcommand{\ee}{\end{equation}}

\newcommand{\no}{\nonumber\\}
\newcommand{\ba}{\begin{eqnarray}}
\newcommand{\ea}{\end{eqnarray}}
\newcommand{\ci}[1]{\cite{#1}}
\newcommand{\bi}[1]{\bibitem{#1}}
\newcommand{\la}[1]{\label{#1}}
\def\gl#1{(\ref{#1})}

\title{\bf CP violation in the models of fermion localization on a domain wall (brane)}
\author{
Alexander A. Andrianov$^{ab}$, Vladimir A. Andrianov$^{a}$, Oleg O. Novikov$^{a}$
\\
$^{a}$ V.A. Fock Department of Theoretical Physics,\\\phantom{$^{a}$} Sankt-Petersburg State University,\\
\phantom{$^{a}$}
ul. Ulianovskaya, 198504 St. Petersburg, Russia\\
$^{b}$ Institut de Ci\`encies del Cosmos, Universitat de Barcelona,\\ \phantom{$^{a}$}
Marti i Franques, 1, 08028
Barcelona, Spain
\\
{\small E-mail: {\tt andrianov@icc.ub.edu, v.andriano@rambler.ru, oonovikov@gmail.com}}}

\date{}

\begin{document}

\maketitle
\begin{abstract} A brief survey of fermion localization mechanism on a domain wall (“thick brane”) generated by a topologically nontrivial vacuum configuration of scalar fields is given. The extension  of scalar fields interaction with fermions which supplies fermions  with an axial mass is proposed.
For several flavors and generations of fermions  this extension can entail the Cabibbo-Kobayashi-Maskawa matrix of the Standard Model. As well the model with two scalar doublets which provide a supplementary CP violation mechanism is considered.
\end{abstract}

\section{Introduction}

The hypothesis that our universe is a four-dimensional space-time hypersurface (3-brane) embedded in a fundamental multi-dimensional space \ci{rushap,otherbr} has become popular basis for models of Beyond the Standard Model physics \ci{ADD}-\ci{RSII}, see, for example, the reviews \ci{RuBar}-\ci{loc6} and references therein. It is assumed that the extra dimensions size is large enough and they can be, in principle, detected in terrestrial experiments planned in the near future and/or in astrophysical observations.

Brane is often considered as an elementary geometrical object of a vanishing thickness. However there is an alternative provided by an effective multi-dimensional field theory \ci{rushap}. The brane in this approach is a domain wall generated by background scalar and/or gravitational fields \ci{rev121}-\ci{rev19} when their vacuum configuration has a non-trivial topology (Section 3). The details of the localization of the matter on this domain walls may provide important clues to the low energy physics. Among other possibilities it can play an important role in fermion mass generation \cite{loc1}-\cite{loc5} (see Section 2). In this paper we examine possible ways to introduce CP violation to an effective four-dimensional theory by means of fermion localization mechanism. In Section 4 the generalization of the model elaborated earlier in \ci{aags1} with extra Yukawa vertices to provide fermions with complex mass is presented. In the case of several flavors it allows to construct a mass matrix that can be associated with the Cabibbo-Kobayashi-Maskawa (CKM) matrix of the Standard Model. In Section 5 we also present the model with two scalar doublets that may provide a new source of CP violation.

\section{Fermion localization on a domain wall}

Let us start with elucidating how to trap a fermion matter on a four-dimensional hyperplane, i.e. the domain wall ("thick brane").
The extra-dimension coordinate is assumed to be space-like,

$$
(X_\alpha) = (x_\mu, y)\ , \quad (x_\mu) = (x_0, x_1, x_2, x_3)\ , \quad
(\eta_{\alpha\alpha}) = (+,-,-,-,-)
$$
while the subspace $x_\mu$ corresponds to the four-dimensional Minkowski space. The extra-dimension size
is assumed to be infinite (or large enough). In this paper we neglect the gravity and assume the bulk space-time
to be flat. The fermion wave function obeys the Dirac equation,
\be
[\,i\gamma_\alpha \partial^\alpha - \Phi(X)\,]\psi(X) = 0\ , \quad
\gamma_\alpha = (\gamma_\mu, -i\gamma_5)\ ,\quad \{\gamma_\alpha,
\gamma_\beta\}
= 2\eta_{\alpha\beta}\ , \la{5dir}
\ee
with  $\gamma_\alpha$ being a set of four-dimensional Dirac matrices in the
chiral representation. The trapping of light fermion states on a four-dimensional
hyperplane -- the domain wall -- the "thick brane" can be provided by a topologically
nontrivial background configuration of the scalar field,
 $\langle\Phi(X)\rangle_0 = \varphi (y)$,
which provides the appearance of zero-modes in the four-dimensional
fermion spectrum. For the four-dimensional space-time interpretation,
\gl{5dir} can be decomposed into the infinite set of fermions with
different masses,
\ba
&&[\,i\gamma_\alpha \partial^\alpha + \varphi(y)\,][\,i\gamma_\alpha
\partial^\alpha - \varphi(y)\,]\psi(X)
\equiv (- \partial_\mu \partial^\mu - \widehat m^2_y) \psi(X)\ ;\no
&&\widehat m^2_y = - \partial_y^2 + \varphi^2 (y) - \gamma_5 \varphi'(y) =
\widehat m^2_{+} P_L + \widehat m^2_{-} P_R\ , \la{f-s}
\ea
where  $P_{L,R} = \frac12 (1 \pm \gamma_5)$ are projectors on
the left- and right-handed states.
Thus the mass squared operator $\widehat m^2_y$ consists of two chiral
 partners,
\ba
\widehat m_\pm^2 &=&  - \partial_y^2 + \varphi^2 (y) \mp \varphi'(y)
=  [\,-\partial_y \pm \varphi(y)\,][\,\partial_y \pm \varphi(y)\,]\ ;
\la{fact}\\
\widehat m_{+}^2\,q^+ &=& q^+\,\widehat m_{-}^2,\quad
\widehat m_{-}^2\,q^- = q^-\,\widehat m_{+}^2\ ,\quad q^\pm \equiv
\mp \partial_y + \varphi(y)\ . \la{susy}
\ea
Due to such a supersymmetry \ci{susy1,susy2}, for non-vanishing masses,
 the left- and right-handed spinors in \eqref{susy} form the bi-spinor
describing a dim-4 massive Dirac particle which is, in general, not
localized at any point of the extra-dimension for asymptotically constant
field configurations $\varphi(y)$. Such a spectral equivalence may be broken
by a normalizable zero mode of
one of the mass operators $\widehat m_\pm^2$. This mode is read out of
 Eqs.\ \gl{fact} and \gl{susy},
\be
q^-\psi^{+}_0(x,y) = 0\ , \quad \psi^{+}_0(x,y) =
\psi_L(x) \, \exp\left\{-\int^y_{y_0} dw\varphi(w)\right\}\ ,
\ee
where $\psi_L(x) = P_L \psi (x)$ is a free-particle
Weyl spinor in the four-dimensional Minkowski space.
Evidently, if a scalar field configuration has
the appropriate asymptotic behavior,
$$
\varphi(y)\stackrel{y \rightarrow \pm\infty}{\sim}
\pm C_\pm |y|^{\nu_\pm}\ ,\quad \mbox{\rm Re} \nu_\pm > -1\ ,
\quad  C_\pm > 0\ ,
$$
then the wave function $\psi^{+}_0(x,y)$ is normalizable
on the $y$ axis and the corresponding
left-handed fermion is a massless Weyl particle localized in the vicinity
of a four-dimensional domain wall.
The example of  an appropriate topological configuration  is realized by
a kink-like scalar field background (of possibly dynamical origin, see below),
\be
\varphi^{+} =
M\, \mbox{\rm tanh}(My)\ . \la{soli}
\ee
The two mass operators have the following potentials,
\be
\widehat m^2_{+} =- \partial_y^2 + M^2
\left[\,1-2{\rm sech}^2(My)\,\right];\quad
\widehat m^2_{-} =- \partial_y^2 + M^2, \la{chpot}
\ee
and the left-handed normalized zero-mode is localized around $y=0$,
\be
\psi^{+}_0(x,y) =
\psi_L(x)\,\psi_0 (y)\ ,\qquad \psi_0 (y) \equiv
\sqrt{M/2}\ {\rm sech}(My)\ . \la{locmod}
\ee
Evidently
the threshold for the continuum is at $ M^2$
and the heavy Dirac particles may have
any masses $m > M$. The corresponding
wave functions are  spread out in the fifth dimension.
For light fermions
on a brane one needs
at least two five-dimensional fermions $\psi_1(X), \psi_2(X)$
in order to generate left- and right-handed
parts of a four-dimensional Dirac bi-spinor as zero modes.
The required zero modes with different chiralities for
$\langle\Phi(X)\rangle_0 = \varphi^+(y)$ arise when the two fermions couple
to the  scalar field $\Phi(X)$ with opposite charges,
\be
[\,i\not\!\partial - \tau_3\Phi(X)\,]\Psi(X) = 0\ ,\quad
\not\!\partial \equiv \widehat\gamma_\alpha \partial^\alpha\ ,\quad
\Psi(X) =\left\lgroup\begin{array}{c}\psi_1(X)\\
\psi_2(X)\end{array}\right\rgroup\ ,
\la{2fer}
\ee
where $\widehat\gamma_\alpha \equiv \gamma_\alpha\otimes {\bf 1}_2$ are
Dirac matrices and
$\tau_a \equiv {\bf 1}_4 \otimes \sigma_a,\ a=1,2,3 $
are the generalizations of
the Pauli matrices $\sigma_a$ acting on the bi-spinor components $\psi_i(X)$.
The next task
is to supply this massless particle with a light mass. As the mass operator
mixes left- and right-handed
components of the four-dimensional fermion it is embedded in the Dirac
operator \gl{2fer} with the mixing matrix $\tau_1 m_f$ of the fields
$\psi_1(X)$ and $\psi_2(X)$.
If realizing the Standard Model mechanism
of fermion mass generation by means of dedicated scalars,
one has to introduce the
second scalar field $H_1(X)$, replacing the bare mass
$\tau_1 m_f \longrightarrow \tau_1 H_1(X)$ in the lagrangian density
 \cite{aags1}. In this paper we generalize this mechanism to produce
axial fermion mass by introducing extra term $\tau_2 H_2(X)$,
\ba
{\cal L}^{(5)} (\bar\Psi,\Psi,\Phi, H) =
\bar\Psi ( i\!\not\!\partial
- \tau_3 \Phi -\tau_1 H_1 - \tau_2 H_2) \Psi . \la{aux}
\ea
Note that arbitrary $\tau_3$ rotation
 $\Psi\rightarrow \exp\{i\frac{\alpha}{2}\tau_3\}\Psi$
 corresponds to the chiral
transformation of the localized four-dimensional fermions
  $\psi_0 (x,y)\rightarrow \exp\{i\frac{\alpha}{2}\gamma_5\}\psi_0 (x,y)$.
This transformation is equivalent to the rotation,
\be
H_1\rightarrow H_1\cos{\alpha}+H_2\sin{\alpha},\quad H_2\rightarrow H_2\cos{\alpha}-H_1\sin{\alpha}.
\ee
The phase $\alpha$ corresponds to the arbitrary phase factor that can be added to
 fermion masses and Yukawa constants.

\section{The formation of a brane in the model with two scalar fields}
In this Section we study the brane formation  in the model with two scalar fields with a
 potential admitting kink solutions. The model that was studied in detail in \ci{aags1} possesses a quartic $O(2)$ symmetric
 self-interaction as well as quadratic terms for both scalar fields providing
  soft breaking of $O(2)$ symmetry,
\be
{\cal L}_{scal}^{(5)}=Z\Bigl(\frac{1}{2}\partial_\alpha\Phi\partial^\alpha\Phi+
\frac{1}{2}\partial_\alpha H\partial^\alpha H+M^2\Phi^2+\Delta_HH^2-
\frac{1}{2}\Bigl(\Phi^2+H^2\Bigr)^2\Bigr).
\label{ScalarLagrangian}
\ee
The normalization coefficient $Z$ has dimension of a mass and is introduced to
simplify the equation structure.

We restrict ourselves to the background solutions not breaking 4D Lorentz
invariance. The classical Eqs. of motion take the form,
\ba
\Phi''=-2M^2\Phi+2(\Phi^2+H^2)\Phi,\\
H''=-2\Delta_HH+2(\Phi^2+H^2)H,\label{ScalarEq}
\ea
Depending on the relation between the coupling constants $M^2$ and $\Delta_H$ there
exist two solutions of Eqs. (\ref{ScalarEq}) inhomogeneous in $y$ which
correspond to two phases: the phase with vanishing v.e.v. of $H$ and the phase with nonzero v.e.v. of $H$,
with the second-order phase transition at $\Delta_H=M^2/2$. The second phase under our interest corresponds
to
$M^2/2 \leq \Delta_H \leq M^2$, i.e.
 $2\Delta_H = M^2 +\mu^2, \ \mu^2 < M^2$, when
\be
\Phi(y)  = \pm M\tanh \left( {\beta  y} \right), \quad H (y) =
 \pm \frac{\mu}{\cosh \left( {\beta  y} \right)} ,\quad \beta =
  \sqrt{M^2 - \mu^2}. \label{zeroap} \ee
The second variation operator for scalar fields $\Phi$ and $H$ can be written as,
\be
\mathcal{D}_X^2=-\square_x-\mathcal{M}^2,
\ee
where
\be
(\mathcal{M}^2)_{ij}=\delta_{ij}\Bigl[-\partial_y^2-2\Delta_i+2S_kS_k\Bigr]+
4S_i S_j,\; S=\begin{pmatrix}\Phi(y)\\H(y)\end{pmatrix},\Delta=\begin{pmatrix}M^2\\
\Delta_H\end{pmatrix}
\ee
is a mass operator that determines the spectrum of the scalar fluctuations $\Omega_\phi(x,y)$ and $\Omega_h(x,y)$.

If taking $\mu/M$ as a small perturbation parameter which
controls the deviation from the critical point $\mu=0$ the mass
 operator can be rewritten in the approximate form \cite{aags1},
\ba
& (\hat{\mathcal{M}}^2)_{11}\approx (-\partial_y+2M\tanh(My))
(\partial_y+2M\tanh(My))+4\mu^2\tanh^2(M y),\no
& (\hat{\mathcal{M}}^2)_{22}\approx (-\partial_y+M\tanh(My))
(\partial_y+M\tanh(My))+4\mu^2{\rm sech}^2(M y),\no
& (\hat{\mathcal{M}}^2)_{12}=(\hat{\mathcal{M}}^2)_{21}\approx 4\mu M\tanh(M y){\rm sech}(M y)
\ea
The scalar fluctuation spectrum consists of two localized states and the continuous spectrum
of delocalized states with masses above $2M$. The first localized state is a massless
Goldstone mode associated with translation symmetry breaking and can
be obtained exactly by differentiating the background fields $\Phi(y)$ and $H(y)$ with respect to $y$ ,
\be
\Omega_\phi=\sqrt{\frac{3\beta}{2(2M^2+\mu^2)Z}}\begin{pmatrix}
{M\rm sech}^2(\beta  y)\\-\mu\tanh(\beta y){\rm sech}(\beta y)\end{pmatrix}
\phi(x),\quad\partial^\mu\partial_\mu \phi=0.
\ee
The second state describes a light scalar particle. At the critical point it is a zero-mode however for
non-zero $\mu$ it gains a mass and its wave function can be written as,
\ba
&\Omega_h=
\sqrt{\frac{M}{2Z}}\begin{pmatrix}-\mu y{\rm sech}^2(My)\\{\rm sech}(My)
\end{pmatrix}h(x)+O\Bigl(\frac{\mu^2}{M^2}\Bigr),
\quad\partial^\mu\partial_\mu h=m_h^2h,\no
&m_h^2=2\mu^2\Bigl(1+O\Bigl(\frac{\mu^2}{M^2}\Bigr)\Bigr).
\ea
Notice that we neglected influence of gravity (its influence on the background
fields was studied in \cite{aags2,aao}). It was shown earlier (see for example \ci{rev15,rev19}) that
 a mixing of the scalar and graviscalar degrees of freedom may change the
spectrum in the physical scalar sector drastically. This leads to the absence
of the translational Goldstone mode. On the other
hand the  mass and the profile of wave function for light scalar state remain the same at least in
the leading order. Because at low energies the Goldstone mode is decoupled
and fermions interact only with the light massive scalar we assume that the effects nonperturbative in gravitational constant do not make any influence on the results
presented further on.

\section{Generation of complex mass for localized fermions}
Let us consider the interaction of two five-dimensional fermions to the  background scalar doublet described in the previous section,
\be
\mathcal{L}_{f}=\begin{pmatrix}\Psi_1\\\Psi_2\end{pmatrix}^{\dagger}\gamma^0
\Bigl(i\gamma_\alpha \partial^\alpha -g_A\Phi\tau_3-g_1H\tau_1-g_2H\tau_2\Bigr)
\begin{pmatrix}\Psi_1\\\Psi_2\end{pmatrix} .
\ee
To study the four-dimensional physics we decompose the fields into the
infinite set of fermions with certain 4D mass,
\be
\begin{pmatrix}\Psi_1\\\Psi_2\end{pmatrix}=\sum\limits_m\begin{pmatrix}
F^{(m)}_{1L}(y)\\F^{(m)}_{2L}(y)\end{pmatrix}\psi^{(m)}_L(x)+\begin{pmatrix}
F^{(m)}_{1R}(y)\\F^{(m)}_{2R}(y)\end{pmatrix}\psi^{(m)}_R(x).
\ee
The profile functions of variable $y$ can be represented as the superposition of the two
sets of solutions,
\be
F_{1L}\equiv F_L,\, F_{1R}\equiv F_R,\quad F_{2L}=\mp F_R^{\ast},\,
F_{2R}=\pm F_L^{\ast} .
\ee
These two solutions can be related to each other by the chiral
 transformation $\psi\rightarrow\gamma^5\psi$. In general, changing of the profile
   functions phase is equivalent to the chiral rotation of fermions.

Eqs. on these profile functions read,
\be
(\partial_y+g_A\Phi)F_L-(g_1-ig_2)HF_R^{\ast}=m_fF_R;
 \label{FermionEq1Left}
\ee
\be
(-\partial_y+g_A\Phi)F_R+(g_1-ig_2)HF_L^{\ast}=m_f^{\ast}F_L .
 \label{FermionEq1Right}
\ee

Assuming that the parameter $\mu/M\equiv\epsilon\ll 1$ we can  treat the
 term with $H$ as a perturbation,
\be
(\partial_y+g_A\Phi)F_L^{(0)}=(m_f^{(0)})^{\ast}F_R^{(0)},\quad(-\partial_y
+g_A\Phi)F_R^{(0)}=m_f^{(0)}F_L^{(0)}.
\ee
Then at zero order in $\epsilon$ there is no a localized solution with nonzero $F_R$,
\be
F_L^{(0)}=N{\rm sech}^{\frac{g_AM}{\beta}}(\beta y),\quad F_R^{(0)}=0,\quad
 m_f^{(0)}=0;
\ee
\be
N=\sqrt{\frac{\beta\Gamma(p+\frac{1}{2})}{\sqrt{\pi}\Gamma(p)}},
\quad p\equiv\frac{g_AM}{\beta} ,
\ee
where $\Gamma(p)$ is a gamma function.
It is easy to see that $F_L$ is an even function while $F_R$ is odd. Multiplying
 \eqref{FermionEq1Right} by $(F_L^{(0)})^\ast$, integrating it from
  $-\infty$ to $+\infty$ and assuming that both $F_L$ and $F_R$ are localized we
   obtain the leading-order fermion mass,
\be
m_f^{(1)}=
(g_1+ig_2)\frac{\int_{-\infty}^{+\infty}dy|F_L^{(0)}|^2H}{
\int_{-\infty}^{+\infty}dy|F_L^{(0)}|^2}=
(g_1+ig_2)\mu\frac{\int_{-\infty}^{+\infty}dy\,{\rm sech}^{2p+1}(\beta y)}{
\int_{-\infty}^{+\infty}dy\,{\rm sech}^{2p}(\beta y)}=\nonumber
\ee
\be
=(g_1+ig_2)\mu\frac{\left(
\Gamma(p+\frac{1}{2})\right)^2}{\Gamma(p)\Gamma(p+1)}.
\ee
For the physics on a brane at energies much lower than $M$ only the lightest
 localized states are relevant. The effective lagrangian up to quadratic order
  in scalar field fluctuations takes the form,
\ba
&\mathcal{L}_{low}=\frac{1}{2}\partial_\mu\phi\partial^\mu\phi+\frac{1}{2}
\partial_\mu h\partial^\mu h-\frac{m_h^2}{2}h^2+\no
&+i\bar{\psi}\gamma^\mu\partial_\mu\psi-
\left[\psi_R^\dagger(m_f+g_{\phi}\phi+g_{h}h)\psi_L+h.c.\right] .
\ea
The Yukawa constant for Goldstone boson $\phi$ vanishes from symmetry
 considerations,
\be
g_{\phi}=
\frac{\int_{-\infty}^{+\infty}dy\left(2F_R^{\ast}F_L\Omega_{\phi,\Phi}
+(|F_L|^2+|F_R|^2)\Omega_{\phi,H}\right)}{
\int_{-\infty}^{+\infty}dy(|F_L|^2+|F_R|^2)}=0.
\ee
The light scalar state plays the  role of a Higgs-like boson with the Yukawa
constant,
\be
g_{h}=
\frac{\int_{-\infty}^{+\infty}dy\left(2F_R^{\ast}F_L\Omega_{h,\Phi}+
(|F_L|^2+|F_R|^2)\Omega_{h,H}\right)}{
\int_{-\infty}^{+\infty}dy(|F_L|^2+|F_R|^2)}=\nonumber
\ee
\be
=\sqrt{\frac{M}{2Z}}(g_1+ig_2)
\frac{\left(
\Gamma(p+\frac{1}{2})\right)^2}{\Gamma(p)\Gamma(p+1)}
+O(\epsilon^2)=\sqrt{\frac{M}{Z}}\frac{m_f}{m_h},
\ee
where $\epsilon \equiv \mu/M$ .
In the case of one flavor the phase factor of both the mass and the Yukawa constant can
 be  always removed by the chiral transformation
  $\psi\rightarrow e^{i\theta\gamma_5}\psi$.
 This model can be easily
   generalized to include several flavors,
\be
\mathcal{L}_{f}=\sum\limits_{m,n = 1}^{n_f}
\begin{pmatrix}\Psi_{m,1}\\\Psi_{m,2}\end{pmatrix}^{\dagger}
\gamma^0\Bigl(i\gamma_\alpha\partial^\alpha \delta_{mn}-g_{mn,A}\Phi\tau_3-g_{mn,1}H\tau_1-
g_{mn,2}H\tau_2\Bigr)\begin{pmatrix}\Psi_{n,1}\\\Psi_{n,2}\end{pmatrix} .
\ee
Because the coupling constants admit mixing of different flavors we have to
 introduce the mass matrix $M_{mn}$ which can be connected with the CKM matrix
  of the Standard Model. Eqs. on profile functions read,
\be
\sum\limits_{n = 1}^{n_f}\left[(\partial_y\delta_{mn}+g_{mn,A}\Phi)F_{n,L}-(g_{mn,1}-ig_{mn,2})HF_{n,R}^{\ast}\right]=\sum\limits_{n = 1}^{n_f}
M_{mn}F_{n,R} ;\label{FermionEqMLeft}
\ee
\be
\sum\limits_{n = 1}^{n_f}\left[(-\partial_y\delta_{mn}+g_{mn,A}\Phi)F_{n,R}+(g_{mn,1}-ig_{mn,2})HF_{n,L}^{\ast}\right]=\sum\limits_{n = 1}^{n_f}
(M^\dagger)_{mn}F_{n,L} ,\label{FermionEqMRight}
\ee
and the low energy lagrangian becomes as follows,
\ba
&\mathcal{L}_{low}=\frac{1}{2}\partial_\mu\phi\partial^\mu\phi+
\frac{1}{2}\partial_\mu h\partial^\mu h-\frac{m_h^2}{2}h^2+\no
&+i\sum\limits_{m = 1}^{n_f}\bar{\psi}_m\gamma^\mu\partial_\mu\psi_m - \sum\limits_{m,n = 1}^{n_f}
\left[\psi_{mR}^\dagger(M_{mn}+g_{mn,\phi}\phi+g_{mn,h}h)\psi_{nL}+
h.c.\right].
\ea
If the Yukawa constant for the kink $\Phi(y)$ is universal $g_{mn,A}=g_A$ then
 $F_{n,L}^{(0)}$ is the same for all flavors and there is a simple relation
  between the leading orders for the mass matrix and Yukawa constants,
\be
M_{mn}\simeq(g_{mn,1}+ig_{mn,2})\mu\frac{\left(
\Gamma(p+\frac{1}{2})\right)^2}{\Gamma(p)\Gamma(p+1)},\quad g_{mn,h}
\simeq\sqrt{\frac{M}{Z}}\frac{M_{mn}}{m_h}.\label{MassYukawaMatrix}
\ee

\section{Model with two scalar doublets and CP-violation}
Let us consider now a model with two independent  scalar
 doublets with v.e.v.'s described in Sec.3. We restrict ourselves by the
  following toy-model with particular fermion couplings,
\ba
&\mathcal{L}_5=
\begin{pmatrix}\Psi_1\\\Psi_2\end{pmatrix}^{\dagger}\gamma^0
\Bigl(i\gamma_\alpha\partial^\alpha-(g_{1A}\Phi_1+g_{2A}\Phi_2)\tau_3-g_1H_1\tau_1
-g_2H_2\tau_2\Bigr)\begin{pmatrix}\Psi_1\\\Psi_2\end{pmatrix}+\no
&+Z_1\Bigl(\frac{1}{2}\partial_\alpha\Phi_1\partial^\alpha\Phi_1+\frac{1}{2}\partial_\alpha
 H_1\partial^\alpha H_1+M_1^2\Phi_1^2+\Delta_1H_1^2-
 \frac{1}{2}\Bigl(\Phi_1^2+H_1^2\Bigr)^2\Bigr)+\no
&+Z_2\Bigl(\frac{1}{2}\partial_\alpha\Phi_2\partial^\alpha\Phi_2+
\frac{1}{2}\partial_\alpha H_2\partial^\alpha H_2+M_2^2\Phi_2^2+\Delta_2H_2^2
-\frac{1}{2}\Bigl(\Phi_2^2+H_2^2\Bigr)^2\Bigr) .
\ea
Because the scalar doublets are independent they can form kinks at different
 positions $y-a$ and $y+a$,
\ba
&\Phi_1=M_1\tanh{\beta_1(y-a)},\quad H_1=\mu_1{\rm sech}{\beta_1(y-a)},\\
&\Phi_2=M_2\tanh{\beta_2(y+a)},\quad H_2=\mu_2{\rm sech}{\beta_2(y+a)} .
\ea
Note that different values of the couplings (including the Yukawa constants)
 lead to the asymmetry of fermion profiles. For simplicity we assume that $\beta_1=\beta_2\equiv\beta$
   and $\mu_1/M_1,\mu_2/M_2,\beta a\sim \epsilon\ll 1$.

The equations on the fermion profile functions take the form,
\be
(\partial_y+g_{1A}\Phi_1+g_{2A}\Phi_2)F_L-(g_1H_1-ig_2H_2)F_R^\ast=m_fF_R;
 \label{FermionEq2Left}
\ee
\be
(-\partial_y+g_{1A}\Phi_1+g_{2A}\Phi_2)F_R+(g_1H_1-ig_2H_2)F_L^\ast=
m_f^{\ast}F_L. \label{FermionEq2Right}
\ee
The calculation similar to what was done in the previous Section yields the
 following result,
\be
F_L=\sqrt{\frac{\beta\Gamma(\tilde{p}+\frac{1}{2})}{\sqrt{\pi}
\Gamma(\tilde{p})}}{\rm sech}^{\tilde{p}}(\beta y)+O(\epsilon),
\quad F_R^{(0)}=O(\epsilon),
\ee
\be
m_f=(g_1\mu_1+ig_2\mu_2)\frac{\left(
\Gamma(\tilde{p}+\frac{1}{2})\right)^2}{\Gamma(\tilde{p})\Gamma(\tilde{p}+1)}
\cdot\Bigl(1+O(\epsilon)\Bigr)=|m_f|e^{i\theta},
\ee
\be
\tilde{p}\equiv\frac{g_{1A}M_1+g_{2A}M_2}{\beta},\quad \tan\theta=
\frac{g_2m_{h2}}{g_1m_{h1}}.
\ee
The scalar sector at low energies includes two massless Goldstone bosons and two
 light scalars,
\ba
&\mathcal{L}_{low}=
\frac{1}{2}\partial_\mu\phi_1\partial^\mu\phi_1+
\frac{1}{2}\partial_\mu h_1\partial^\mu h_1-\frac{m_{h1}^2}{2}h_1^2+\no
&+\frac{1}{2}\partial_\mu\phi_2\partial^\mu\phi_2+\frac{1}{2}\partial_\mu h_2
\partial^\mu h_2-\frac{m_{h2}^2}{2}h_2^2+\no
&+i\bar{\psi}\gamma^\mu\partial_\mu\psi-
\left[\psi_R^\dagger(m_f+g_{\phi1}\phi_1+g_{\phi2}\phi_2+g_{h1}h_1+
g_{h2}h_2)\psi_L+h.c.\right].
\ea
The Goldstone bosons decouple from fermions in the lowest orders in $\epsilon$ at least up to
 $O(\epsilon^3)$. The leading order of light-scalar Yukawa couplings takes the
  following form,
\be
g_{h1}\simeq\cos\theta\sqrt{\frac{M_1}{Z_1}}\frac{|m_f|}{m_{h1}},
\quad g_{h2}\simeq i\sin\theta\sqrt{\frac{M_2}{Z_2}}\frac{|m_f|}{m_{h2}}.
\ee
While the phase factor of the fermion mass can be removed by the chiral
transformation
 $\psi\rightarrow \exp\Bigl(-i\gamma_5\frac{\theta}{2}\Bigr)\psi$
  the Yukawa coupling constants remain complex providing
  the new source of CP violation in addition to the complex phase factor
    of the CKM matrix,
\ba
&\tilde{g}_{h1}\simeq
(\cos^2\theta-i\sin\theta\cos\theta)
\sqrt{\frac{M_1}{Z_1}}\frac{|m_f|}{m_{h1}},\\
&\tilde{g}_{h2}\simeq
 (\sin^2\theta+i\sin\theta\cos\theta)
 \sqrt{\frac{M_2}{Z_2}}\frac{|m_f|}{m_{h2}}.
\ea

\section{Discussion}

We have presented the mechanisms of inducing CP violation in some models of fermions
 localized on a domain wall ("thick brane"). The model with one scalar doublet mostly
  reproduces the Standard Model. Note however that the simple relation
 \eqref{MassYukawaMatrix} between coupling constants and masses holds only in the leading order of $\mu/M$ when
 different flavors have the same profile function. A more substantial deviation
 from the Standard Model may come from the non-universality of the $g_A$
 constant. The model with two scalar doublets provides an extra source of CP
 violation from different phases of two Higgs-like scalars Yukawa
 constants. This paper was focused only on fermion localization
 mechanism. The gauge boson couplings and experimental implications
 of these models are left aside for further investigation.

\medskip

{\bf Acknowledgements. }
This work was done under financial support by RFBR, project No. 13-02-00127 and Saint-Petersburg State University grant 11.38.660.2013. A.A.Andrianov was also supported by projects FPA2010-20807 and CPAN (Consolider
CSD2007-00042).

\end{document}